\documentclass{iau} 
\usepackage{graphicx}
\usepackage{amsmath,bm}	
\usepackage{amssymb}

\def\MC{meridional circulation}
\def\vb{{\bf v}}
\def\Fn{{\bf F}_{\nu}}
\def\FL{{\bf F}_L}

\title[Variation of Meridional Circulation with solar cycle] 
{Explaining the variation of the meridional circulation with the solar cycle}

\author[Hazra \& Choudhuri]   
{Gopal Hazra
 \and Arnab Rai Choudhuri}

\affiliation{Dept. of Physics, Indian Institute of Science, \\Bangalore - 560012,
India \\ email: {\tt hgopal, arnab@iisc.ac.in}}

\pubyear{2018}
\volume{340}  
\jname{Long-Term Datasets for the Understanding of Solar and Stellar Magnetic Cycles}
\editors{Dipankar Banerjee, Jie Jiang, Kanya Kusano \& Sami Solanki}
\begin{document}

\maketitle

\begin{abstract}
The meridional circulation of the Sun is observationally found to vary with the solar cycle, becoming
slower during the solar maxima. We explain this by 
constructing a theoretical model in which
the equation of the meridional circulation (the $\phi$ component of the vorticity equation) 
is coupled with the equations of the flux transport dynamo model. We
find that the Lorentz force of the dynamo-generated magnetic fields can slow down the \MC\ during
the solar maxima in broad conformity with the observations.

\end{abstract}

\firstsection
\section{Introduction}
The meridional circulation is one of the most important large-scale coherent flow
patterns within the solar convection zone. 
It has been found from helioseismic measurements that the
meridional circulation varies with the solar cycle
(Chou \& Dai 2001; Zhao \& Kosovichev 2004; Komm {\em et al.} 2015). These results are consistent with
the surface measurements of Hathaway \& Rightmire (2010),
who found that the \MC\ at the surface becomes weaker during the sunspot
maximum by an amount of order 5 m s$^{-1}$. We explore whether this
variation of the \MC\ can be due to the Lorentz force 
of the dynamo-generated magnetic field.

The \MC\ plays a very critical role in the flux transport dynamo
model, which started being developed from the mid-1990s (Choudhuri, Sch\"ussler
\& Dikpati 1995; Durney 1995). Most of the papers on the flux transport dynamo are of kinematic
nature, which do not include the back-reactions of the magnetic field
on the large-scale flows. In order to include such back-reactions, we have
to couple the mean field dynamo equations with the dynamical equation governing
the flow (i.e. essentially the Navier--Stokes equation). This is what we do in our
recent paper (Hazra \& Choudhuri 2017). The next two Sections summarize the
methodology and the results presented by Hazra \& Choudhuri (2017).

There is some observational evidence for random fluctuations in the \MC\ having
a time scale longer than the solar cycle.  These fluctuations seem crucial in producing such
effects as the Waldmeier effect (Karak \& Choudhuri 2011), as well as the observed
correlation of the decay rate of a cycle with the strength of the next cycle (Hazra {\em et al.}
2015). These fluctuations also make a contribution in producing grand minima (Choudhuri
\& Karak 2012). In this study, we do not include such fluctuations and consider only
dynamo-induced periodic variations. Although there are indications that the spatial
structure of the \MC\ is more complicated than the single-cell pattern assumed
in many dynamo models (see Hazra, Karak \& Choudhuri 2014), 
it is still a debated issue and we assume a single-cell structure.

\section{Methodology}
To incorporate the effect of the Lorentz force on the meridional circulation, we need to consider the $\phi$ component of the vorticity equation, which is
\begin{eqnarray}
\label{eq:main1}
\frac{\partial \omega_{\phi}}{\partial t} + s \nabla.\left(\vb_m \frac{\omega_{\phi}}{s}\right) =  
s \frac{\partial \Omega^2}{\partial z}+ \frac{1}{\rho^2}(\nabla \rho \times \nabla p)_{\phi} \\ ~~\nonumber
+ [\nabla \times \FL]_{\phi} + [\nabla \times \Fn(\vb_m)]_{\phi}.
\end{eqnarray} 
where $\omega_{\phi}$ is the $\phi$ component of vorticity which comes from the 
meridional circulation $\vb_m$ only, $s = r\sin \theta$, $\FL$ is the Lorentz
force term and $\Fn (\vb_m)$ is the turbulent viscosity term corresponding to the velocity
field $\vb_m$. We now break up the the meridional velocity into two parts:
\begin{equation}\label{eq:v_tot}
 \vb_m = \vb_0 + \vb_1,
\end{equation}
where $\vb_0$ is the regular meridional circulation the Sun would have in the absence of
magnetic fields and $\vb_1$ is its modification
due to the Lorentz force of the dynamo-generated magnetic field. The azimuthal vorticity $\omega_{\phi}$
can also be broken into two parts corresponding to these two parts of the \MC:
\begin{equation}\label{eq:vort}
 \omega_{\phi} = \omega_0 + \omega_1.
\end{equation}

We substitute Eqs.~\ref{eq:v_tot} and \ref{eq:vort} in Eq.~\ref{eq:main1}. Then we subtract from 
it the version of Eq.~\ref{eq:main1} for $\vb_0$ alone (${\bf F}_L$ will be absent in this case). Neglecting 
quadratic terms in perturbed quantities, we get
\begin{eqnarray}
\label{eq:main2}
\frac{\partial \omega_1}{\partial t} + s \nabla. \left(\vb_0 \frac{\omega_1}{s}\right) + 
s \nabla. \left( \vb_1 \frac{\omega_0}{s}\right) \\ ~\nonumber
= [\nabla \times \FL]_{\phi} + [\nabla \times \Fn(\vb_1)]_{\phi}
\end{eqnarray} 

This equation is solved along with the mean field equations of the flux transport dynamo, which give
the magnetic field from which the Lorentz force appearing in Eq.~\ref{eq:main2} is calculated.
A solution of Eq.~\ref{eq:main2} first yields the perturbed vorticity $\omega_1$
at different steps. We can then compute the perturbed velocity $\vb_1$ using streamfunction-vorticity formalism.
For more details of the methodology, we refer the reader to Hazra \& Choudhuri (2017).
We have assumed that the dynamo-generated magnetic fields do not
affect the thermodynamics significantly, making the thermal wind term $(\frac{1}{\rho^2}(\nabla \rho \times \nabla p)_{\phi})$ to drop out of Eq.~\ref{eq:main2}.
This is what makes the theory of the modification of the \MC\ decoupled from the
thermodynamics of the Sun and simpler to handle than the theory of the unperturbed \MC.


\begin{figure}
\centering{
\includegraphics[width=0.74\textwidth]{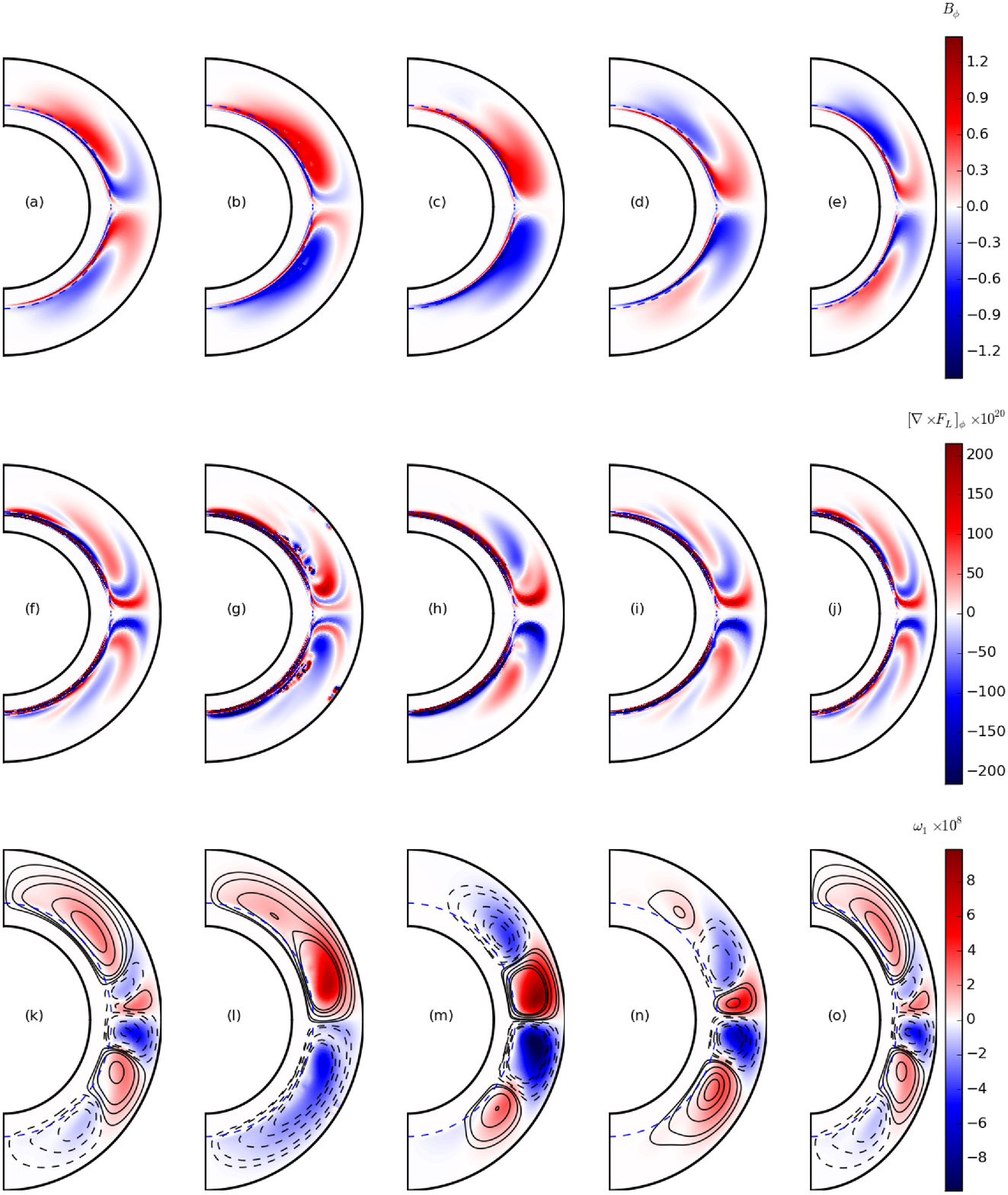}}
\caption{Snapshots in the $r-\theta$ plane of different quantities spanning an entire solar cycle: the toroidal field
$B_{\phi}$, the $\phi$ component of the curl of the Lorentz force $[\nabla \times \FL]_{\phi}$
and the perturbed vorticity $\omega_1$ with streamlines are shown in the first row (a)-(e), the second row (f)-(j), and 
the third row (k)-(o) respectively. The five columns represent time instants during
the midst of the rising phase, the solar maximum, the midst of the decay phase and the solar minimum,
followed by the midst of the rising phase again. In the third row (k)-(o), the solid black contours represent clockwise flows and the dashed black contours 
represent anti-clockwise flows. Taken from Hazra \& Choudhuri (2017).} 
\label{fig:snap_1d12}
\end{figure}

\section{Results}

Our discussions in this Section will refer to the northern hemisphere of the Sun, some quantities having
the opposite sign in the southern hemisphere.
The unperturbed meridional circulation there is anti-clockwise, implying
a negative vorticity $\omega_0$. We need to generate a positive perturbed vorticity
$\omega_1$ at the time of the sunspot maximum, if the meridional circulation is to be weakened by the Lorentz
force at that time. Figure~\ref{fig:snap_1d12} reproduced from Hazra \& Choudhuri (2017) illustrates this.  The three rows
of this figure plot three quantities in the $r-\theta$ plane: the toroidal field $B_\phi$, the 
$\phi$-component of the curl of the Lorentz force $[\nabla \times {\bf F}_L]_{\phi}$ and the perturbed
vorticity $\omega_1$ along with the associated streamlines. The five vertical columns correspond to five
time intervals during a solar cycle.  The second column corresponds to a time close to the sunspot
maximum, whereas the the fourth column corresponds to the sunspot minimum. 

It is clear from the second column of Figure~\ref{fig:snap_1d12} (corresponding to the solar 
maximum) that $[\nabla \times {\bf F}_L]_{\phi}$ is negative in the northern hemisphere at
that time, giving rise to a negative perturbed vorticity. This will clearly make the 
\MC\ weaker at that time.  When the flow velocity is calculated from the total vorticity
as given by Eq.~\ref{eq:vort}, we find that it varies in a periodic fashion. Figure~2 shows
the time evolution of the total $v_{\theta}$ at $25^{\circ}$ latitude just below the surface
along with the sunspot number.  We see that the \MC\ reaches its minimum a little after the
sunspot maximum. This figure can be compared with observed variation of the meridional flow at the 
surface (Fig.~4 of Hathaway \& Rightmire 2010).

\begin{figure}[!h]
\centering{
\includegraphics[width=0.65\textwidth]{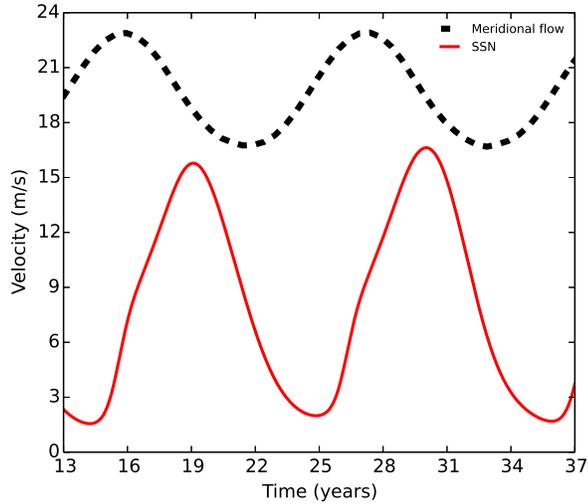}}
\caption{The black dashed line shows the time variation of the total meridional circulation just below the surface at $25^{\circ}$ latitude. The red solid line shows the yearly averaged sunspot number. Taken from Hazra \& Choudhuri (2017).}
\label{fig:tot_mc}
\end{figure}

It may be noted that, while solving the dynamo equations in this exploratory study, we have
used only the unperturbed \MC\ given by $\vb_0$. If we want to include the full time-varying \MC\
in the dynamo equations, then we have to solve the streamfunction-vorticity equation at every time
step to find the perturbed velocity from the perturbed vorticity. Although this is computationally expensive,
we plan to do this in future for a typical illustrative case. While a varying \MC\ would affect the
magnetic field generated by the dynamo (Karak \& Choudhuri 2011; Choudhuri \& Karak 2012), a \MC\
varying periodically with the solar cycle is not expected to change the behavior of the system
qualitatively (Karak \& Choudhuri 2012).

\section{Conclusion}

We conclude that the back-reaction due to the Lorentz force of the dynamo-generated magnetic field
on the meridional circulation is the most likely reason for its variation with the solar cycle and 
our results are in good agreement with the observed variation of the meridional circulation.

{\it Acknowledgments.} A.R.\ Choudhuri's
research is supported by DST through 
a J.C.\ Bose Fellowship.

\end{document}